\documentclass[twocolumn,superscriptaddress, pre]{revtex4-2}
\usepackage{amsmath,amssymb}
\usepackage{graphicx}
\usepackage{dcolumn}
\usepackage{bm}
\usepackage{xr}

\makeatletter
\newcommand*{\addFileDependency}[1]{
  \typeout{(#1)}
  \@addtofilelist{#1}
  \IfFileExists{#1}{}{\typeout{No file #1.}}
}
\makeatother

\newcommand*{\myexternaldocument}[1]{%
    \externaldocument{#1}%
    \addFileDependency{#1.tex}%
    \addFileDependency{#1.aux}%
}

\myexternaldocument{supp}

\newcommand{\diffusivity}{\bm{D}}
\newcommand{\force}{f_\mathrm{chem}}

\newcommand{\vmax}{v_\mathrm{max}}
\newcommand{\Dc}{D_\mathrm{c}}
\newcommand{\Dm}{D_\mathrm{m}}
\newcommand{\xmi}{x_{i}}

\newcommand{\xc}{x_\mathrm{c}}
\newcommand{\dotxmi}{\dot{x}_{i }}

\newcommand{\etaS}{\eta_\mathrm{S}}

\newcommand{\CV}{\theta}
\newcommand{\x}{\bm{x}}
\newcommand{\type}{Letter} 
\newcommand{\Dbare}{D_\mathrm{bare}}
\newcommand{\mean}{\bm{\mu}}
\newcommand{\cov}{\bm{C}}
\newcommand{\A}{\bm{A}}
\newcommand{\lin}{\bm{u}}
\newcommand{\const}{\bm{v}}

\begin{document}

\title{Dynamic and Thermodynamic Bounds for Collective Motor-Driven Transport}

\author{Matthew P. Leighton}
\email{matthew\_leighton@sfu.ca}
\author{David A.\ Sivak}%
\email{dsivak@sfu.ca}
\affiliation{Department of Physics, Simon Fraser University, Burnaby, British Columbia, V5A 1S6, Canada.}%

\date{\today}

\begin{abstract}
Molecular motors work collectively to transport cargo within cells, with anywhere from one to several hundred motors towing a single cargo. For a broad class of collective-transport systems, we use tools from stochastic thermodynamics to derive a new lower bound for the entropy production rate which is tighter than the second law. This implies new bounds on the velocity, efficiency, and precision of general transport systems and a set of analytic Pareto frontiers for identical motors. In a specific model, we identify conditions for saturation of these Pareto frontiers.
\end{abstract}

\maketitle

\emph{Introduction}.---Molecular transport motors like kinesin and myosin are constantly at work within the cells of every living organism on Earth, consuming chemical energy in order to accomplish important tasks~\cite{howard2001mechanics}. Their roles include transporting molecular cargo against concentration gradients~\cite{chaffey2003alberts} and applying directed forces to facilitate cell division~\cite{hatsumi1992mutants} or contract muscle tissue~\cite{cooke1997actomyosin}. Motor-driven transport systems \emph{in vivo} consist of many coupled subsystems moving together: cargo such as vesicles~\cite{encalada2011stable}, organelles~\cite{hancock2008intracellular}, or actin filaments~\cite{cooke1997actomyosin} pulled by anywhere from only one~\cite{shtridelman2008force} to several hundred~\cite{leopold1992association} motor proteins.

Many specific models of transport systems have been explored, including deterministic phenomenological models~\cite{klumpp2005cooperative,shtridelman2008force,shtridelman2009vivo}, discrete stochastic models~\cite{bhat2016transport,bhat2017stall,brown2019pulling}, and continuous stochastic models~\cite{mckinley2012asymptotic,leighton2022performance}. A common goal of these investigations has been to determine how various parameters (such as coupling strength, stall force, diffusivity, and number of motors) tune the performance of these systems. Performance metrics of interest include dynamical quantities such as velocity and precision and thermodynamic quantities like efficiency and power consumption~\cite{brown2019theory}. While the behavior of specific model systems has been explored, considerably less is known about the fundamental performance limits for transport systems in general, agnostic of model details.

The behavior of transport systems is restricted by two fundamental thermodynamic limitations. First and foremost, they must obey the second law of thermodynamics, the most useful form in these contexts stating that at steady state the ensemble-averaged rate of global entropy production cannot be negative~\cite{seifert2012stochastic}. Second, the recently established thermodynamic uncertainty relation (TUR)~\cite{barato2015thermodynamic,gingrich2016dissipation,horowitz2020thermodynamic} lower bounds products of the entropy production rate and uncertainties in various currents at steady state. These key inequalities have been used to derive bounds on various performance metrics, for example efficiency~\cite{pietzonka2016universal}.

In this work we consider the thermodynamics of motor-driven intracellular transport, where a coupled collection of active and passive components travel together at steady state. We show that these systems obey a new bound, derived from Jensen's inequality, on their total entropy production. This bound is always tighter than the second law, and is often tighter than the TUR (and much easier to estimate). From this Jensen bound and the TUR, we derive a set of bounds on performance metrics such as velocity, efficiency, and precision. These bounds constrain emergent properties of collective systems for arbitrary number of motors of any directionality, using only bare properties of individual subsystems. Our theory holds for a broad class of collective-transport systems, independent of any model-specific interaction potentials or spatially inhomogeneous energy landscapes. For identical motors, we then derive analytic expressions for several Pareto frontiers constraining combinations of performance metrics. Finally, we simulate an example system to illustrate these bounds and conditions sufficient for their saturation.

\emph{Theory and model}.---Consider $N$ transport motors coupled to a diffusing molecular cargo, all moving in one dimension. Each motor interacts with the cargo via a molecular linker, and is characterized by a mechanochemical cycle through which it transduces chemical power into directed forward motion. The cargo undergoes overdamped Brownian motion (with bare diffusivity $\Dc$) constrained by interactions with each motor (with its own bare diffusivity $D_i$).

Each motor in isolation experiences a constant chemical driving force $f_i$, along with a spatially periodic potential-energy landscape arising due to interactions with the substrate it walks along (e.g., microtubules for kinesin). This may include features such as metastable states and energy barriers. (Multiple cargos are trivially incorporated as motors with no chemical driving force, $f_i=0$.) Motors and cargo are coupled via the total potential energy $V(\bm{x})$ for $\bm{x} \equiv \{\xc,x_1,...,x_i,...,x_N\}$ the vector of cargo position $\xc$ and motor positions $\{\xmi\}_{i=1}^N$. This potential describes, e.g., the molecular linkers attaching each motor to the cargo and attractive or repulsive interactions between motors.

In the long-time limit the subsystems (cargo and motors) must stay together; i.e., the relative coordinates $x_i-\xc$ reach time-independent distributions at steady state so that each subsystem has the same mean velocity, $\langle v\rangle \equiv \langle \dot{x}_\mathrm{c}\rangle = \langle \dot{x}_i\rangle$. In terms of the potential $V(\x)$, this requires that all subsystems are coupled and at long distances any repulsive interactions are dominated by attraction. 

The system dynamics are assumed multipartite~\cite{horowitz2015multipartite},meaning that each subsystem (cargo and each motor) is subject to independent thermal fluctuations, so that the system's probability distribution evolves according to the Fokker-Planck equation~\cite{risken1996fokker}
\begin{equation}
\begin{aligned}
\partial_t P(\bm{x},t) & = -\partial_{\xc} J_\mathrm{c}(\bm{x},t)
- \sum_{i=1}^N \partial_{\xmi}J_i(\bm{x},t).
\end{aligned}
\end{equation}
Here $\partial_{x_\alpha}$ is the partial derivative with respect to $x_\alpha$, and the subsystem probability currents are
\begin{subequations}
\begin{align}
J_\mathrm{c}(\bm{x},t) & = \Dc\left[ -\beta \frac{\partial V}{\partial \xc} -  \frac{\partial}{\partial \xc}\right]P(\bm{x},t)\\
J_i(\bm{x},t) & = D_i \left[\beta f_i -\beta \frac{\partial V}{\partial \xmi}-  \frac{\partial}{\partial \xmi}\right]P(\bm{x},t).
\end{align}
\end{subequations}
Here $\beta = (k_\mathrm{B}T)^{-1}$ is the inverse temperature.

We focus on system behavior at its ``steady state'': the limiting regime in which time evolution is independent of initial conditions. Mathematically, the relevant limit is that the time $t\gg\tau_\mathrm{relax}$, the system's longest relaxation time. We assume that this limit exists, and that steady-state properties such as velocity, efficiency, energy flows, and entropy production all have well-defined constant averages.

The mean velocity is an integral over the probability current for each subsystem~\cite{seifert2012stochastic}:
\begin{subequations}\label{eq:vAv}
\begin{align}
\langle v\rangle & = \left\langle \frac{J_\mathrm{c}(\bm{x},t)}{P(\bm{x},t)}\right\rangle\\
& = \left\langle \frac{J_i(\bm{x},t)}{P(\bm{x},t)}\right\rangle.
\end{align}
\end{subequations}
Angle brackets denote ensemble averages. 

Each subsystem exchanges heat with the thermal reservoir at temperature $T$, and each motor exchanges chemical energy with a chemical reservoir at constant chemical potential. Likewise, motors and cargo exchange energy with each other through their interaction potentials. Of particular interest is the average rate of total chemical-energy consumption by the $N$ motors:
\begin{subequations}
\label{eq:Pchem}
\begin{align}
P_\mathrm{chem} & \equiv \sum_{i=1}^N \langle f_i \dotxmi \rangle\\
& =  \sum_{i=1}^N f_i\langle v\rangle.
\end{align}
\end{subequations}
This definition implicitly assumes each motor tightly couples its chemical and mechanical degrees of freedom, consistent with experiments on kinesin and myosin motors~\cite{schnitzer1997kinesin,visscher1999single,toyoshima1990myosin}.

Transport systems do not in general store energy, so their thermodynamic efficiency is zero. A natural (and positive) measure of their efficiency is the Stokes efficiency~\cite{wang2002stokes},
\begin{equation}\label{stokes_def}
    \etaS \equiv \frac{\zeta_\mathrm{c} \langle v\rangle^2}{P_\mathrm{chem}}
\end{equation}
that quantifies the fraction of the consumed chemical energy that produces work pulling the cargo against viscous friction, characterized by friction coefficient $\zeta_\mathrm{c}=1/(\beta\Dc)$.

The above metrics [Eqs.~\ref{eq:vAv}-\ref{stokes_def}] quantify the average behavior of transport systems; we quantify long-time stochasticity by the effective diffusivity
\begin{equation}\label{eq:D_eff}
D_\mathrm{eff} \equiv \lim_{t\to\infty} \frac{\langle \delta \xc^2\rangle}{2t},
\end{equation}
and precision by the coefficient of variation
\begin{equation}\label{CV_def}
\CV \equiv \frac{\sqrt{\langle \delta \xc^2\rangle}}{\langle x_\mathrm{c}\rangle},
\end{equation}
with $\langle \delta \xc^2\rangle$ the variance of the cargo position $\xc$. Reference~\cite{leighton2022performance} evaluates and discusses the above metrics in a specific example system.

The average rates of dimensionless entropy production for each subsystem are~\cite{horowitz2015multipartite}
\begin{subequations}\label{entropy_production}
\begin{align}
\dot{\Sigma}_{i} & = \frac{1}{D_i}\left\langle \left[\frac{J_i(\bm{x},t)}{P(\bm{x},t)}\right]^2\right\rangle \geq 0\\
\dot{\Sigma}_\mathrm{c} & = \frac{1}{\Dc}\left\langle \left[\frac{J_\mathrm{c}(\bm{x},t)}{P(\bm{x},t)}\right]^2\right\rangle \geq 0.
\end{align}
\end{subequations}
The total entropy production rate is their sum, $\dot{\Sigma} = \dot{\Sigma}_\mathrm{c} + \sum_{i=1}^N\dot{\Sigma}_{i}$. For a diffusive cargo with no external forces, the entropy production equals the total chemical power:
\begin{equation}\label{entropy_power_equation}
\dot{\Sigma} = \beta P_\mathrm{chem}.
\end{equation}

\emph{Bounds for general systems}.---Given the functional form of the average velocity~\eqref{eq:vAv}, Jensen's inequality~\cite{cover1999elements} requires 
\begin{equation}
\langle v\rangle^2 \leq  \left\langle \left[\frac{J_
\alpha(\bm{x},t)}{P(\bm{x},t)}\right]^2\right\rangle.
\end{equation}
Three inequalities follow from this, constraining the partial and total entropy production rates:
\begin{subequations}\label{entropy_inequalities}
\begin{align}
\dot{\Sigma}_{i}& \geq \frac{1}{D_i}\langle v\rangle^2,\\
\dot{\Sigma}_\mathrm{c}& \geq \frac{1}{\Dc}\langle v\rangle^2,\\
\dot{\Sigma} & \geq \frac{1}{\Dbare}\langle v\rangle^2. \label{main_entropy_inequality}
\end{align}
\end{subequations}
Here $\Dbare$ is the ``bare collective diffusivity,'' the inverse of the total friction coefficient from summing the individual friction coefficients (inversely proportional to bare diffusivities) of each subsystem:
\begin{equation}
\Dbare \equiv \left(\frac{1}{\Dc} + \sum_{i=1}^N \frac{1}{D_i}\right)^{-1}.
\end{equation}
Physically, $\Dbare$ is the effective diffusivity under a potential that only depends on relative subsystem positions.

This ``Jensen bound'' [Eq.~\eqref{main_entropy_inequality}] is our first major result: a general, model-independent, lower bound [non-negative and thus tighter than the second law~\eqref{entropy_production}] on the entropy production required for a collective-transport system with $N$ motors to maintain mean velocity $\langle v\rangle$. 

The collective-transport system is also constrained by the long-time limit of the thermodynamic uncertainty relation~\cite{barato2015thermodynamic,gingrich2016dissipation,horowitz2020thermodynamic}, whose most useful form for this system is 
\begin{equation}\label{TUR}
\dot{\Sigma}t\,\frac{\langle\delta \xc^2\rangle}{\langle \xc\rangle^2}\geq 2.
\end{equation}
Identifying $\langle v\rangle = \langle \xc\rangle/t$ and $D_\mathrm{eff}$~\eqref{eq:D_eff} recasts this inequality as
\begin{equation}\label{TURIneq}
\dot{\Sigma}\geq \frac{1}{D_\mathrm{eff}}\langle v\rangle^2, 
\end{equation}
which has the same form as our Jensen bound Eq.~\eqref{main_entropy_inequality}. Equations~\eqref{main_entropy_inequality} and \eqref{TURIneq} thus constitute two bounds on the entropy production. In general, either of these bounds can be tighter. Even for a single particle in a tilted sinusoidal potential, either $\Dbare<D_\mathrm{eff}$ or $\Dbare>D_\mathrm{eff}$ is possible, depending on the ratio of the barrier height to the driving force~\cite{reimann2001giant}.

Substituting Eq.~\eqref{entropy_power_equation} and the Stokes efficiency~\eqref{stokes_def} into Eq.~\eqref{main_entropy_inequality} gives an upper bound on $\eta_\mathrm{S}$:
\begin{equation}\label{eff_ineq}
\etaS \leq \frac{\Dbare}{\Dc}.
\end{equation}
This is similar, but not equivalent, to a previous bound~\cite{pietzonka2016universal}: $\eta_\mathrm{S}\leq D_\mathrm{eff}/\Dc$. Like the Jensen bound~\eqref{main_entropy_inequality} and TUR~\eqref{TURIneq}, either of these two bounds can be tighter in different circumstances.

Likewise, substituting Eq.~\eqref{entropy_power_equation} and $P_\mathrm{chem} = f_\mathrm{tot}\langle v\rangle $ (for total force $f_\mathrm{tot} = \sum_{i=1}^N f_i$ 
which we assume without loss of generality to be non-negative)
into Eq.~\eqref{main_entropy_inequality} yields an upper bound on the average velocity:
\begin{equation}\label{vel_ineq}
\langle v\rangle \leq \beta \Dbare f_\mathrm{tot}.
\end{equation}

Finally, substituting Eq.~\eqref{entropy_power_equation} and the coefficient of variation Eq.~\eqref{CV_def} into the TUR~\eqref{TUR} and employing the velocity inequality~\eqref{vel_ineq} gives an upper bound on the precision through a lower bound on the coefficient of variation:
\begin{equation}\label{CV_ineq}
\CV \geq \frac{1}{\beta f_\mathrm{tot}}\sqrt{\frac{2}{\Dbare t}}
\end{equation}

These three bounds [Eqs.~\eqref{eff_ineq}-\eqref{CV_ineq}] constitute our second major result, constraining global system properties using only properties ($\Dbare$, $\Dc$, and $f_\mathrm{tot}$) of each individual subsystem in isolation.

\emph{Identical motors}.---We illustrate the utility of these performance bounds with the special case where transport motors are identical, each with diffusivity $\Dm$ and driving force $f_\mathrm{chem}$. This reflects many biological systems of interest, such as identical kinesin motors towing a large vesicle, or identical myosin motors pulling an actin filament. The Jensen bound~\eqref{main_entropy_inequality} becomes
\begin{equation}\label{identical_entropy_ineq}
\dot{\Sigma} \geq \left(\frac{1}{\Dc} + \frac{N}{\Dm}\right)\langle v\rangle^2.
\end{equation}

Our general performance bounds [Eqs.~\eqref{eff_ineq}-\eqref{CV_ineq}] can be rewritten in terms of more natural variables as
\begin{subequations}
\begin{align}
\etaS & \leq \left(1 + \frac{N\Dc}{\Dm}\right)^{-1},\label{eff_ineq_identical}\\
\langle v\rangle & \leq \beta N \Dc \force \left(1 + \frac{N\Dc}{\Dm}\right)^{-1},\label{vel_ineq_identical}\\
\CV & \geq \frac{1}{\beta N\force}\sqrt{\frac{2}{\Dc t}}\left(1 + \frac{N\Dc}{\Dm}\right)^{1/2}.
\end{align}
\end{subequations}
Since $N\Dc/\Dm>0$, a looser upper bound on the mean velocity is 
\begin{equation}\label{eq:vmaxdef}
v_\mathrm{max} = \beta \Dm \force,
\end{equation}
the mean velocity of a single motor in a flat potential subject to constant force $\force$. Likewise, since $N\geq1$, the Stokes efficiency has a looser upper bound of $\left(1+ \Dc/\Dm\right)^{-1}$.

Combining Eq.~\eqref{eff_ineq_identical} with Eq.~\eqref{vel_ineq_identical} gives a Pareto frontier between the Stokes efficiency and scaled mean velocity:
\begin{equation}\label{pareto_v_eta}
\etaS + \frac{\langle v\rangle}{v_\mathrm{max}}\leq 1.
\end{equation}
Similarly, combining Eqs.~\eqref{identical_entropy_ineq} and \eqref{vel_ineq} gives
\begin{equation}\label{PV_frontier}
\frac{P_\mathrm{chem}}{P_{\to i}^\mathrm{max}} \geq \frac{\Dm}{\Dc}\frac{\left(\langle v\rangle/v_\mathrm{max}\right)^2}{1 - \langle v\rangle/v_\mathrm{max}},
\end{equation}
a Pareto frontier constraining velocity and power consumption. Here $P_{\to i}^\mathrm{max} = \force\vmax$ is the mean power consumption of a single motor at maximum velocity. These two Pareto frontiers follow solely from the Jensen bound~\eqref{main_entropy_inequality}; the TUR~\eqref{TUR} alone gives a Pareto frontier for power consumption and precision:
\begin{equation}
\beta P_\mathrm{chem}\theta^2\geq2.
\end{equation}

So far the cargo has only encountered resistance from viscous drag; similar considerations also constrain performance for an additional external force $f_\mathrm{ext}$ on the cargo, in the direction opposite to $\force$. The entropy production rate is then
\begin{equation}
\begin{aligned}
\dot{\Sigma} & = \beta(N\force - f_\mathrm{ext})\langle v\rangle\\
& \geq \left(\frac{1}{\Dc} + \frac{N}{\Dm}\right)\langle v\rangle^2,
\end{aligned}
\end{equation}
Here thermodynamic efficiency $\eta_\mathrm{T} \equiv f_\mathrm{ext}/(N\force)$ is positive. Applying the Jensen bound leads to a Pareto frontier for thermodynamic efficiency and mean velocity: 
\begin{equation}
\eta_\mathrm{T} + \frac{\Dm}{N\Dbare}\frac{\langle v\rangle}{v_\mathrm{max}}\leq1.
\end{equation} 
Since $\Dm/\,N\Dbare\geq1$, a looser bound 
analogous to Eq.~\eqref{pareto_v_eta} is 
\begin{equation}\label{thermeffineq}
\eta_\mathrm{T} 
+ \frac{\langle v\rangle}{v_\mathrm{max}}\leq 1.
\end{equation}

\emph{Example system}.---Consider an example with tunable parameters that can saturate our derived bounds. Each motor has periodic potential $V_i(\xmi) = \frac{1}{2}E^\ddagger \cos \left(2\pi\xmi/\ell\right)$ with barrier height $E^\ddagger$, period $\ell$, and maximum conservative force $f_\mathrm{max} = E^\ddagger/(2\ell)$. Each motor is linked to the cargo by a Hookean spring with spring constant $\kappa$ and zero rest length~\cite{kawaguchi2003equilibrium}, $U_i(\xc,\xmi) = \frac{1}{2} \kappa (\xmi-\xc)^2$. The motors do not directly interact. The total system potential is thus
\begin{equation}\label{eq:total_potential}
V(\x) = \sum_{i=1}^N \left[\frac{1}{2}E^\ddagger \cos \left(2\pi\xmi/\ell\right) + \frac{1}{2} \kappa (\xmi-\xc)^2\right].
\end{equation}

\begin{figure}[t!]
\includegraphics[width=\columnwidth]{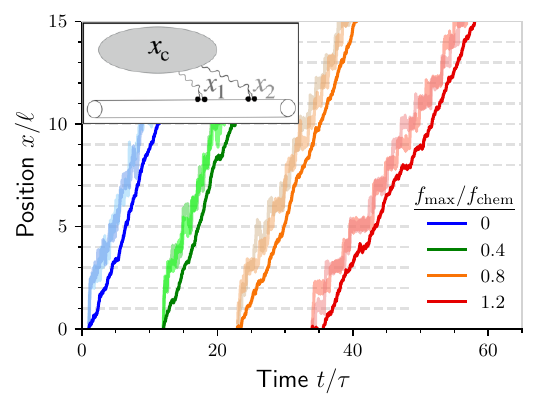} 
\caption{\label{fig:Dynamics} Motor and cargo trajectories for the example system with $N=2$ motors, for different $f_\mathrm{max}/\force$. Dark curves, cargo; lighter curves, two motors. Dashed gray horizontal lines show local minima of motor potential energy. The start times of different trajectories are staggered for clarity. Position and time are, respectively, scaled by $\ell$ and $\tau = \ell^2/\Dm$. Parameters used are $\beta\force\ell = 15$, $\beta\kappa\ell^2 = 7$, and $\Dc/\Dm = 1/30$.}
\end{figure}

Equating the Kramers rate~\cite{kramers1940brownian} for a single uncoupled motor hopping between adjacent landscape minima with experimentally measured rates for kinesin-1 motors~\cite{vu2016discrete} (see Supplemental Material ``Barrier heights in real systems'' for details~\cite{SInote}) yields $f_\mathrm{max}/f_\mathrm{chem} \approx0.4$, which sets the scale of our parameter sweep.

Figure~\ref{fig:Dynamics} illustrates that for $N=2$ motors the dynamics change significantly as the barrier height increases. For $f_\mathrm{max}/f_\mathrm{chem}\ll1$, the motors move continuously, while for $f_\mathrm{max}/f_\mathrm{chem} \gtrsim 1$ the motors hop between distinct  states.

Figure~\ref{fig:EP_Bounds} compares the entropy production rate for the numerical model to the Jensen bound, TUR, and second law. The Jensen bound is generally the tightest constraint for our best estimates of reasonable model parameters in kinesin-vesicle systems.

More generally, the Jensen bound is tighter whenever $D_\mathrm{eff}>\Dbare$. We numerically explore the ratio $D_\mathrm{eff}/\Dbare$ over a 2D region of parameter space in Fig.~S2, finding that $D_\mathrm{eff}>\Dbare$ (the Jensen bound is tighter) over a wide range of coupling strengths and barrier heights. For sufficiently large energy barriers and motor-cargo coupling, however, $D_\mathrm{eff}<\Dbare$ and thus the TUR is tighter. This is consistent with a previous study of coupled Brownian particles diffusing in a single periodic potential~\cite{evstigneev2009interaction}. At high coupling strengths, subsystems can only cross energy barriers simultaneously~\cite{lathouwers2020nonequilibrium}, making forward progress only with much larger fluctuations whose rarity leads to decreased effective diffusivity. Likewise, high energy barriers could lead to phenomena like hindered diffusion, which lowers the effective diffusivity~\cite{muller2013morphogen}. (Recall that any details of interactions with other subsystems or the substrate only affect $D_\mathrm{eff}$, with $\Dbare$ uniquely determined by the diffusion coefficients of the components making up the system.)

\begin{figure}[t]
\includegraphics[width=\columnwidth]{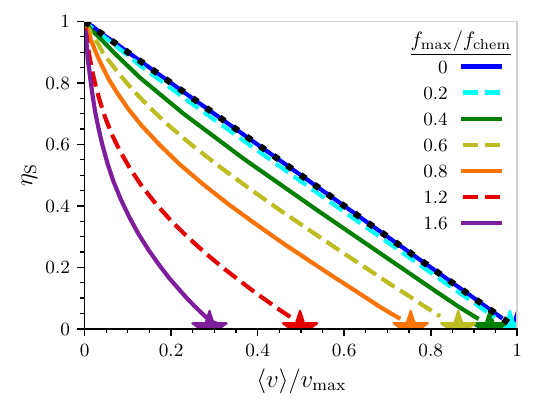} 
\caption{\label{fig:Pareto_Convergence} 
Trade-off between Stokes efficiency $\etaS$ and scaled velocity $\langle v \rangle/v_{\rm max}$ in the example system with $N=2$ motors, plotted parametrically for $\Dc/\Dm =10^{-3}-10^3$. Colors show different $f_\mathrm{max}$. Black dotted line shows the Pareto frontier~\eqref{pareto_v_eta}. Stars show single uncoupled motor. Other parameters same as Fig.~\ref{fig:Dynamics}.}
\end{figure}

Figure~\ref{fig:Pareto_Convergence} shows for $N=2$ motors the trade-off between Stokes efficiency and velocity due to parametric variation of the diffusivity ratio $\Dc/\Dm$, for different barrier heights. When the motors face no barriers ($f_\mathrm{max}/f_\mathrm{chem}=0$), the system exactly saturates the Pareto frontier Eq.~\eqref{pareto_v_eta}. As $f_\mathrm{max}/f_\mathrm{chem}$ increases, the performance trade-off degrades, falling increasingly far from the Pareto frontier.

While the $\etaS$\,--\,$\langle v\rangle$ curve is linear for $f_\mathrm{max}/f_\mathrm{chem}=0$, as $f_\mathrm{max}/f_\mathrm{chem}$ increases it becomes increasingly convex. This suggests that for large energy barriers high efficiency or high velocity are more easily achieved than a compromise between the two. As expected, the velocity in the $N\Dc/\Dm\to\infty$ limit is exactly that of a single uncoupled motor on the same energy landscape, while the Stokes efficiency is zero. In the limit as $N\Dc/\Dm\to0$, the velocity approaches zero and the Stokes efficiency approaches unity.

Beyond this trade-off between efficiency and velocity, the system behaves analogously for other performance trade-offs and metrics; specifically, when $f_\mathrm{max}/\force=0$ this model exactly saturates all our derived bounds. [Figure~\ref{fig:PV_Pareto} illustrates the $P_\mathrm{chem}$\,--\,$\langle v\rangle$ Pareto frontier~\eqref{PV_frontier}].

\emph{Discussion}.---For motor-driven intracellular transport systems, we have derived a new inequality~\eqref{main_entropy_inequality} which lower bounds the entropy production rate of a collective-transport system. This Jensen bound~\eqref{main_entropy_inequality} is always tighter than the second law for a nonstationary transport system, and can be tighter or looser than the thermodynamic uncertainty relation~\eqref{TURIneq}, depending on the relative magnitudes of the bare collective diffusivity $\Dbare$ and the effective diffusivity $D_\mathrm{eff}$. Because of its dependence solely on parameters and averaged quantities, the Jensen bound is much easier to compute than the TUR which depends on $D_\mathrm{eff}$ (a function of the variance, which requires more data to accurately estimate), provided that diffusion coefficients and driving forces are known for each subsystem in isolation.

Once these properties are known for a given set of subsystems, the Jensen bound is easily computed for any collective system assembled from a combination of such modular components. The TUR by contrast does not take advantage of information about the subsystems composing a collective system, and must be computed \textit{de novo} for every such combination by measuring emergent properties of the collective system. This makes the Jensen bound particularly well suited for collective motor-driven transport systems, which are assembled out of parts (cargo and motors) that can be identified and studied in isolation.

Using the Jensen bound and the TUR, we have derived several bounds on performance metrics such as velocity, efficiency, and precision, as well as three analytic expressions for Pareto frontiers when motors are identical. These bounds, which restrict emergent properties of collective systems, depend only on properties of each of arbitrarily many subsystems in isolation. Our results hold quite generally, for arbitrarily many motors (of any directionality) and cargos. The system's joint potential $V(\bm{x})$ is only required to keep the components of the system together at steady state, but may in general capture phenomena not included in our example, such as non-Hookean motor-cargo linkers, motor-motor interactions, or more complex periodic energy landscapes.

Our numerical investigations show that the performance bounds and Pareto frontiers derived in this \type\ are attainable for systems with no energy barriers. This is unsurprising, as it is well known that decreasing energy barriers (catalysis) speeds up a chemical reaction without affecting the energetics. All our bounds and frontiers are saturated for a model with only quadratic couplings between the cargo and each motor. This system, whose dynamics and thermodynamics have been solved analytically~\cite{leighton2022performance}, is Pareto optimal for the class of systems considered here. More generally, the Jensen bound~\eqref{main_entropy_inequality} is always saturated for linear systems within the class of models considered here (see Supplemental Material ``Linear systems saturate the Jensen bound'' for proof~\cite{SInote}). Our simulations focus on $N=2$ motors due to computational constraints; however, our derived bounds hold for arbitrarily large $N$: indeed, their utility is most significant for $N\gg1$, where direct simulation is computationally intractable.

Many of these performance metrics are difficult to measure experimentally, in particular thermodynamic quantities like the chemical power consumption and efficiency; nonetheless, limited experimental measurements of performance trade-offs for \textit{in vivo} systems do exist. Figure~\ref{fig:Experiment} shows measurements of velocity and efficiency for myosin motors in several different animal tissues from Ref.~\cite{purcell2011nucleotide}; for maximum velocity $v_\mathrm{max} = 12\,\mu\mathrm{m/s}$ (to our knowledge, the highest observed in animal muscle tissue~\cite{piazzesi2002size}), our predicted Pareto frontier~\eqref{thermeffineq} indeed bounds the experimentally observed performance. Consistently, theoretical studies of the trade-off between efficiency and velocity in other types of molecular machines have found that high velocity and high efficiency are mutually exclusive~\cite{wagoner2019opposing,wagoner2021evolution}.

While our results apply to a broad class of systems, they do rely on three key assumptions: 1) all components of the transport system stay together, achieving at long times the same mean velocity, 2) the dynamics are multipartite, such that the entropy production can be split into subsystem-specific contributions~\cite{horowitz2015multipartite}, and 3) motor motion is tightly coupled to chemical-energy consumption. Multipartite dynamics are a standard assumption in stochastic thermodynamics~\cite{barato2013information,horowitz2014thermodynamics,horowitz2015multipartite}, generally necessary to analyze the behavior of multicomponent systems. Experiments in kinesin~\cite{schnitzer1997kinesin,visscher1999single} and myosin~\cite{toyoshima1990myosin} motors do support tight coupling between the mechanical and chemical degrees of freedom; nonetheless, futile cycles and backsteps have been observed to occur infrequently~\cite{clancy2011universal}, and are beyond the scope of this \type. We speculate that such phenomena can only degrade the performance metrics discussed in this \type, but generalizing our results to looser mechanochemical coupling will be an important future direction.

\emph{Acknowledgments}.---We thank J.\ Ehrich (SFU Physics) for useful discussions, and J.\ Ehrich and E.\ Jones (SFU Physics) for feedback on the manuscript. We thank the two anonymous reviewers whose comments and suggestions helped improve and clarify this Letter. This work was supported by  Natural Sciences and Engineering Research Council of Canada (NSERC) CGS Masters and Doctoral fellowships (M.P.L.), a BC Graduate Scholarship (M.P.L.), an NSERC Discovery Grant and Discovery Accelerator Supplement (D.A.S.), and a Tier-II Canada Research Chair (D.A.S.).

\bibliography{main}

\appendix
\widetext
\section{Comparison of entropy production bounds}
\begin{figure}[h]
\includegraphics[width=0.8\columnwidth]{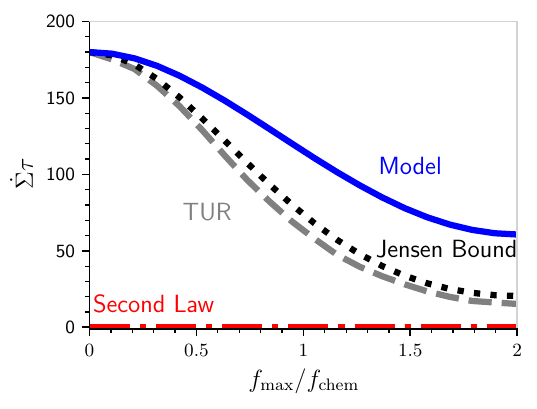}
\caption{\label{fig:EP_Bounds}
Comparison of model entropy production with various lower bounds. Entropy production during time $\tau \equiv \ell^2/\Dm$ of the specific model considered in the main text (blue solid curve), our derived Jensen bound~\eqref{main_entropy_inequality} (black dotted), the TUR~\eqref{TURIneq} (gray dashed), and the second law (red dot-dashed), each as a function of $f_\mathrm{max}/f_\mathrm{chem}$. Uncertainties are smaller than the widths of the curves. Parameters are $N=2$ motors, $\beta\force\ell = 15$, $\beta\kappa\ell^2 = 7$, and $\Dc/\Dm = 1/3$.}
\end{figure}

\begin{figure}[h]
\includegraphics[width=0.8\columnwidth]{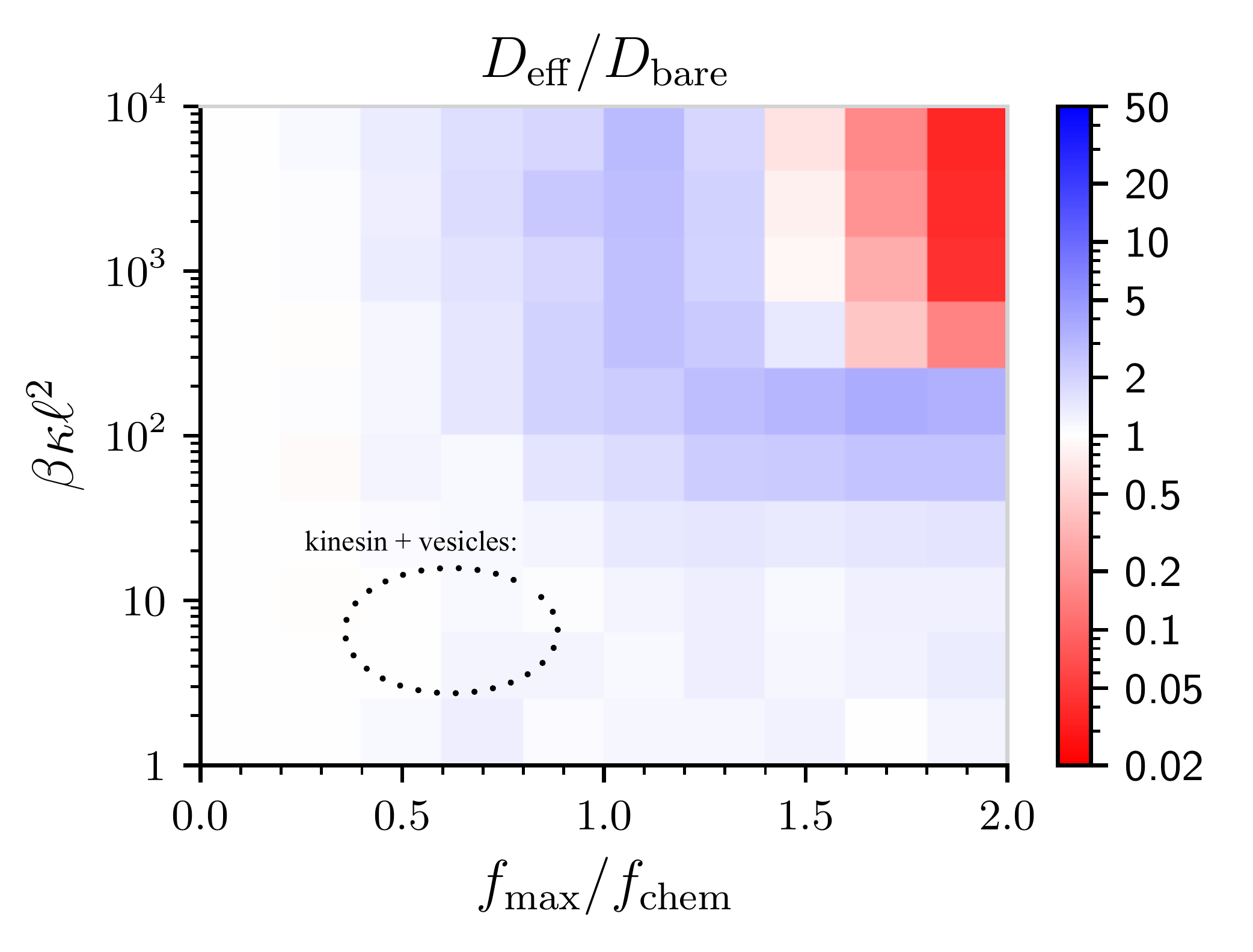}  
\caption{\label{fig:Deff_Dbare} The ratio $D_\mathrm{eff}/D_\mathrm{bare}$ between the effective diffusivity and the bare collective diffusivity, as a function of dimensionless coupling strength $\beta\kappa\ell^2$ and $f_\mathrm{max}/f_\mathrm{chem}$. Parameters are $N=2$ motors, $\beta\force\ell = 15$, and $\Dc/\Dm = 1/3$. Standard errors of the mean are each $\sim$1-5\%.}
\end{figure}

Figure~\ref{fig:EP_Bounds} shows the entropy production rate of the specific model considered in the main text, along with the different lower bounds discussed: the Jensen bound~\eqref{main_entropy_inequality}, TUR~\eqref{TURIneq}, and second law ($\dot{\Sigma}\geq0$). Figure~\ref{fig:Deff_Dbare} shows the ratio $D_\mathrm{eff}/D_\mathrm{bare}$ as a function of coupling strength and barrier height. We expand around our best estimates of $\beta\kappa\ell^2\approx7$ and $f_\mathrm{max}/f_\mathrm{chem}\approx0.4-0.8$ for kinesin motors pulling vesicles. For the most part, $D_\mathrm{eff}>D_\mathrm{bare}$ across the parameter range explored here, except at large barrier heights and high coupling strengths.

\section{Trade-off between velocity and power consumption}
\begin{figure}[h]
\includegraphics[width=0.8\columnwidth]{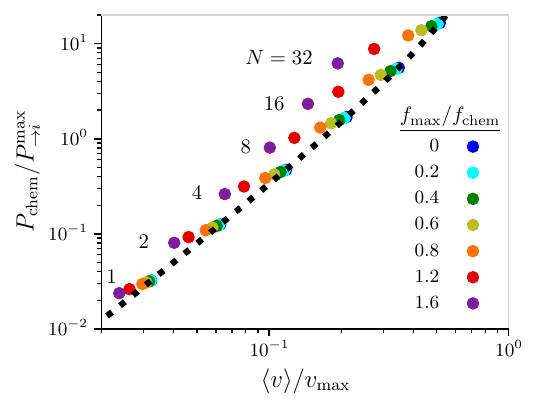}
\caption{\label{fig:PV_Pareto} 
Trade-off between power consumption $P_\mathrm{chem}$ and scaled velocity $\langle v \rangle/v_{\rm max}$ in the example system, plotted parametrically for $N = \{1,2,4,8,16,32\}$ and different $f_\mathrm{max}$ (colors). Black dotted curve: Pareto frontier~\eqref{PV_frontier}. Other parameters same as Fig.~\ref{fig:Dynamics}: $\beta\force\ell = 15$, $\beta\kappa\ell^2 = 7$, and $\Dc/\Dm = 1/30$. Uncertainties are smaller than the widths of the points.}
\end{figure}

Figure~\ref{fig:PV_Pareto} shows the trade-off between power consumption and velocity due to parametric variation of the motor number $N$ and barrier heights. Since this Pareto frontier~\eqref{PV_frontier} depends on $\Dc/\Dm$, we hold that ratio constant. Computational constraints limit us to small $N$. When the motors face no barriers ($f_\mathrm{max}/f_\mathrm{chem}=0$), the system exactly saturates the Pareto frontier~\eqref{PV_frontier}. As $f_\mathrm{max}/f_\mathrm{chem}$ increases, the performance trade-off degrades, falling increasingly far from the Pareto frontier. 

\section{Linear systems saturate the Jensen bound}
Here we prove that a collective-transport system with only linear forces saturates the Jensen bound on entropy production~\eqref{main_entropy_inequality}. Consider a linear system composed of $N+1$ subsystems with positions denoted $\{x_1,...,x_{N+1}\}$, for the first $N$ subsystems the motors, and the last the cargo, so $x_{N+1} \equiv \xc$ and $D_{N+1}\equiv\Dc$. The system has constant force vector $\bm{f}$ and potential 
\begin{equation}
V(\x) = V_0 + \sum_{i=1}^N\sum_{j=i+1}^{N+1} k_{ij}(x_i-x_j)^2.
\end{equation}
We neglect linear terms in the potential since they can be incorporated into the constant forces, and do not allow terms of the form $k_i x_i^2$ that depend on the absolute position of one subsystem, since they preclude the existence of a nonequilibrium steady state. The cargo may in general be subject to a non-zero external force, $f_{N+1} \equiv f_\mathrm{ext}$.

The dynamics of this system are most simply written in Langevin form as
\begin{equation}
\dot{\x} = \beta \diffusivity \left[\bm{f} - \bm{A}\x \right] + \bm{\eta}(t).
\end{equation}
Here $\diffusivity$ is the diffusivity matrix which, under the assumption of multipartite dynamics, is diagonal with entries $D_{ij} = D_i \delta_{ij}$, for Kroneker delta-function $\delta_{ij}$. The matrix $\bm{A}$ satisfies $A_{ij} = \partial_{x_i}\partial_{x_j}V(\x)$, and the vector-valued random noise $\bm{\eta}(t)$ has zero mean and covariance matrix 
\begin{equation}
\left\langle \bm{\eta}(t)\,\bm{\eta}^\top(t')\right\rangle = 2\diffusivity \delta(t-t').
\end{equation}

The solution is a multivariate Gaussian distribution with mean vector $\mean$ and covariance matrix $\cov = \langle \x\x^\top - \mean\mean^\top\rangle$ satisfying the differential equations~\cite[Section 3.2]{risken1996fokker}
\begin{subequations}
\begin{align}
\dot{\mean} & = \beta \diffusivity\left[\bm{f} - \A\mean\right],\\
\dot{\cov} & = -\beta\diffusivity\left[ \A\cov + \A^\top \cov\right] + 2\diffusivity. \label{eq:covode}
\end{align}
\end{subequations}
By definition $\A$ and $\cov$ are symmetric, so $\A=\A^\top$, $\cov=\cov^\top$, and $\cov^{-1} = \left(\cov^{-1}\right)^\top$.

The entropy production rate for the $i$th subsystem is~\cite{horowitz2015multipartite}:
\begin{subequations}
\begin{align}
\dot{\Sigma}_i & = \frac{1}{D_i}\left\langle \left(\frac{J_i(\bm{x},t)}{P(\bm{x},t)}\right)^2\right\rangle\\
& = \frac{1}{D_i} \left\langle \frac{1}{P(\bm{x},t)^2}\left(\beta D_i f_i P(\x,t) - \beta D_i(\A\x)_iP(\x,t) - D_i\frac{\partial}{\partial x_i}P(\x,t)\right)^2\right\rangle\\
& = \frac{1}{D_i} \left\langle\left(\beta D_i f_i - \beta D_i(\A\x)_i - D_i\frac{\partial}{\partial x_i}\ln P(\x,t)\right)^2\right\rangle\\
& = \frac{1}{D_i}\left\langle \underbrace{(\beta D_if_i)^2 - 2(\beta D_i)^2 f_i(\A\x)_i + (\beta D_i(\A\x)_i)^2}_{1} \underbrace{- 2\beta (D_i)^2 \left[f_i - (\A\x)_i\right] \frac{\partial}{\partial x_i}\ln P(\x,t)}_{2} \underbrace{+ D_i^2 \left[\frac{\partial}{\partial x_i}\ln P(\x,t)\right]^2}_{3}\right\rangle. \label{eq:threeTerms}
\end{align}
\end{subequations}
For clarity, we separately evaluate the three terms in this lengthy expression.

The first term is
\begin{subequations}
\begin{align}
 \frac{1}{D_i}\left\langle (\beta D_if_i)^2  - 2(\beta D_i)^2 f_i(\A\x)_i + (\beta D_i(\A\x)_i)^2\right\rangle
& =  \frac{(\beta D_i)^2}{D_i}\left[f_i^2 - 2f_i(\A\mean)_i + \left\langle(\A\x)_i^2\right\rangle\right]\\
& = \frac{(\beta D_i)^2}{D_i}\left[f_i^2 - 2f_i(\A\mean)_i + \left\langle\A\x\x^\top\A\right\rangle_{ii}\right]\\
& = \frac{(\beta D_i)^2}{D_i}\left[f_i^2 - 2f_i(\A\mean)_i + \left(\A\mean\mean^\top\A\right)_{ii} + \left(\A\cov\A\right)_{ii}\right]\\
& = \frac{1}{D_i}\left[(\beta D_i)^2\left(f_i - (\A\mean)_i\right)^2 \right] + \beta^2D_i(\A\cov\A)_{ii}\\
& = \frac{(\dot{\mean}_i)^2}{D_i} + \beta^2D_i(\A\cov\A)_{ii}\label{eq:eqs1}\\
& = \frac{\langle v\rangle^2}{D_i} +\beta D_i\left(\A\right)_{ii}.\label{firstterm}
\end{align}
\end{subequations}
In the last line we took the steady-state limit so that $\dot{\mean}_i = \langle v\rangle$. We further assumed that in the steady-state limit each term in the covariance matrix is linear in $t$:
\begin{equation}
\cov = \lin t + \const,
\end{equation}
where both $\lin$ and $\const$ must be symmetric. This linearity in $t$ is necessary to obtain a constant effective diffusivity in the steady-state limit. The differential equation~\eqref{eq:covode} for the covariance then simplifies to
\begin{subequations}
\begin{align}
\lin & = -2\beta\diffusivity\A\lin t  -2\beta\diffusivity\A\const + 2\diffusivity.
\end{align}
\end{subequations}
Since the left-hand side is independent of $t$, the right-hand side must be as well. For this to be true for general $\diffusivity$, we must have $\A\lin=0$. We then evaluate the rightmost term in \eqref{eq:eqs1}:
\begin{subequations}
\begin{align}
\A\cov\A & = \beta^{-1}\A - \frac{1}{2}\beta^{-1}\diffusivity^{-1}\dot{\cov}\A\\
& = \beta^{-1}\A - \frac{1}{2}\beta^{-1}\diffusivity^{-1}\lin\A\\
& = \beta^{-1}\A - \frac{1}{2}\beta^{-1}\diffusivity^{-1}\left(\A\lin\right)^\top\\
& = \beta^{-1}\A.
\end{align}
\end{subequations}
In \eqref{eq:covode}, the left-hand side is independent of $t$, so the right-hand side must be also, thus $\A\lin = 0.$

The second term in \eqref{eq:threeTerms} is
\begin{subequations}
\begin{align}
\frac{1}{D_i}\left\langle- 2\beta (D_i)^2 \left[f_i - (\A\x)_i\right] \frac{\partial}{\partial x_i}\ln P(\x,t) \right\rangle
& = -2\frac{\beta D_i^2}{D_i}\left\langle \left[f_i - (\A\x)_i\right] \frac{\partial}{\partial x_i}\left(-\frac{1}{2}(\x-\mean)^\top\cov^{-1}(\x-\mean)\right) \right\rangle\\
& = -2\frac{\beta D_i^2}{D_i}\left\langle \left[f_i - (\A\x)_i\right]\left(-\frac{1}{2}\left[(\x-\mean)^\top\cov^{-1}\right]_i - \frac{1}{2}\left[\cov^{-1}(\x-\mean)\right]_i\right) \right\rangle\\
& = 2\beta D_i\left\langle \left[f_i - (\A\x)_i\right] \left( \cov^{-1}(\x-\mean)\right)_{i}\right\rangle\\
& = 2\beta D_i f_i \left\langle \left( \cov^{-1}(\x-\mean)\right)_{i}\right\rangle - 2\beta D_i \left\langle \A\x\left( \cov^{-1}(\x-\mean)\right)_{i}\right\rangle\\
& = -2\beta D_i\left\langle \left(\A\x(\x-\mean)^\top\cov^{-1}\right)_{ii}\right\rangle\\
& = -2\beta D_i\left\langle \left(\A\cov\cov^{-1}\right)_{ii}\right\rangle\\
& = -2\beta D_i\left(\A\right)_{ii}.\label{secondterm}
\end{align}
\end{subequations}

Finally, the third term in \eqref{eq:threeTerms} is
\begin{subequations}
\begin{align}
\frac{1}{D_i}\left\langle D_i^2 \left[\frac{\partial}{\partial x_i}\ln P(\x,t)\right]^2\right\rangle
& = D_i \left\langle \left[-\frac{1}{2}\frac{\partial}{\partial x_i}\left((\x-\mean)^\top\cov^{-1}(\x-\mean)\right)\right]^2\right\rangle\\
& = D_i \left\langle \left( \cov^{-1}(\x-\mean)\right)_{i}^2\right\rangle\\
& = D_i\left\langle \left(\cov^{-1}(\x-\mean)(\x-\mean)^\top\cov^{-1}\right)_{ii}\right\rangle\\
& = D_i \left(\cov^{-1}\left\langle(\x-\mean)(\x-\mean)^\top\right\rangle\cov^{-1}\right)_{ii}\\
& = D_i \left(\cov^{-1}\cov\cov^{-1}\right)_{ii}\\
& = D_i \left(\cov^{-1}\right)_{ii}\\
& = \beta D_i (\A)_{ii}.\label{thirdterm}
\end{align}
\end{subequations} 
To derive the last line we used
\begin{subequations}
\begin{align}
\cov^{-1} & = \A\A^{-1}\cov^{-1}\A^{-1}\A\\
& = \A\left(\A\cov\A\right)^{-1}\A\\
& = \A \left(\beta^{-1}\A\right)^{-1}\A\\
& = \beta\A\A^{-1}\A\\
& = \beta \A.
\end{align}
\end{subequations}

Summing the three terms~\eqref{firstterm}, \eqref{secondterm}, and \eqref{thirdterm}, the entropy production rate of the $i$th subsystem is
\begin{subequations}
\begin{align}
\dot{\Sigma}_i & = \frac{1}{D_i}\langle v\rangle^2 + \beta D_i\left(\A\right)_{ii} - 2\beta D_i\left(\A\right)_{ii} + \beta D_i\left(\A\right)_{ii}\\
& = \frac{1}{D_i}\langle v\rangle^2.
\end{align}
\end{subequations}
Thus the total rate of entropy production is
\begin{equation}
\begin{aligned}
\dot{\Sigma} & = \sum_{i=1}^{N+1}\dot{\Sigma}_i\\
& = \left(\sum_{i=1}^{N+1}\frac{1}{D_i}\right)\langle v\rangle^2\\
& = \left(\frac{1}{\Dc} + \sum_{i=1}^N\frac{1}{D_i}\right)\langle v\rangle^2\\
& = \frac{\langle v\rangle^2}{\Dbare},
\end{aligned}
\end{equation}
exactly saturating the Jensen bound~\eqref{main_entropy_inequality}.

\section{Comparison with experiments}
\begin{figure}[t]
\includegraphics[width=0.8\columnwidth]{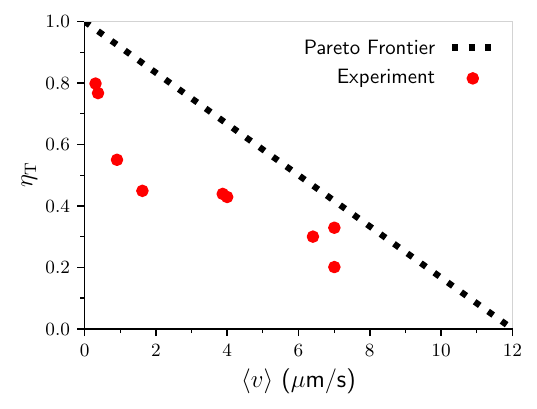}
\caption{\label{fig:Experiment} Myosin motors across various animals obey theoretical Pareto frontier. Red points: experimental measurements of efficiency $\eta_\mathrm{T}$ and velocity $\langle v\rangle$ for myosin motors from different animal species~\cite{purcell2011nucleotide}. Black dotted line: predicted Pareto frontier~\eqref{thermeffineq} for $v_{\rm max} = 12$ $\mu$m/s.}
\end{figure}

Figure~\ref{fig:Experiment} shows experimental measurements of velocity and efficiency for myosin motors in several different animal tissues from Ref.~\cite{purcell2011nucleotide}. For maximum velocity $v_\mathrm{max} = 12\,\mu\mathrm{m/s}$ (to our knowledge, the highest observed in animal muscle tissue~\cite{piazzesi2002size}), our predicted Pareto frontier~\eqref{thermeffineq} indeed bounds the experimentally observed performance. The assumption of a global $v_\mathrm{max}$ across many different species is reasonable so long as the difference between species-specific myosin motors comes predominantly from different potentials $V(\x)$ as opposed to differences in the chemical driving force and bare diffusivity.

\section{Barrier heights in real systems}
Here we use experimental data to estimate the heights of energy barriers separating metastable states for kinesin motors. Recall from \eqref{eq:total_potential} that the $i$th motor has a periodic potential $V_i(\xmi) = \frac{1}{2}E^\ddagger \cos \left(2\pi\xmi/\ell\right)$ with barrier height $E^\ddagger$, period $\ell$, and maximum conservative force $f_\mathrm{max} = E^\ddagger/(2\ell)$. The Kramers rate~\cite{kramers1940brownian} for an uncoupled motor hopping forward from one energy minimum to the next is
\begin{subequations}
\begin{align}
k_+&  = \frac{\beta\Dm}{2\pi}\sqrt{\left|\frac{\partial^2V_i}{\partial x_i^2}\right|_{x_i=a}\cdot \left|\frac{\partial^2V_i}{\partial x_i^2}\right|_{x_i=b}} \,\, e^{-\beta E_b^+}\\
& = \frac{\pi\beta \Dm E^\ddagger}{\ell^2} \, e^{-\beta E_b^+},\label{eq:forwardrate}
\end{align}
\end{subequations}
where $a$ is the position of the bottom of the current potential minimum and $b$ is the position of the peak of the energy barrier to the right. The effective barrier height is $E_b^+ =E^\ddagger - \force\ell/2$. Note that the cosine potential has second derivative of magnitude $2\pi^2E^\ddagger/\ell^2$ at both minima and peaks (points $a$ and $b$).

Likewise, the rate for the motor hopping backward to the previous minimum is
\begin{equation}\label{eq:backwardrate}
k_- = \frac{\pi\beta \Dm E^\ddagger}{\ell^2}\,e^{-\beta E_b^-},
\end{equation}
where this time the effective barrier height is $E_b^- =E^\ddagger + \force\ell/2$.

Analysis of experimental data~\cite{carter2005mechanics} yields step rates for kinesin of $k_+ = 133.0/$s and $k_- = 0.2/$s~\cite{vu2016discrete}, and step size $\ell=8.2$ nm. Combining these with previous estimates of the motor diffusivity $\Dm\approx\mathcal{O}(10^{-3})\,\mu$m$^2$/s~\cite{brown2019pulling,leighton2022performance}, solving the two equations \eqref{eq:forwardrate} and \eqref{eq:backwardrate} for the two remaining parameters yields estimates $\force\ell\approx 7\,k_\mathrm{B}T$ and $E^\ddagger = 2f_\mathrm{max}\ell\approx 6\,k_\mathrm{B}T$. Accordingly, $f_\mathrm{max}/\force\approx0.4$ sets the scale for our numerical investigations.

\end{document}